\documentclass[rmp,aps,preprint,nofootinbib,endfloats]{revtex4}

\usepackage{epsfig}
\usepackage{graphics}

\begin{document}
\title{The Cold Dark Matter crisis  
on galactic and subgalactic scales}
\author{Argyro Tasitsiomi}
\email{iro@oddjob.uchicago.edu}
\affiliation{Department of Astronomy \& Astrophysics, \\
Center for Cosmological Physics,\\
The University of Chicago, \\
5640 S. Ellis  Ave.,\\
Chicago, IL 60637, USA}

\begin{abstract}
The Cold Dark Matter (CDM$/ \Lambda$CDM) model describes  successfully
our Universe on large scales, as has been verified by a wide range of
observations. Nevertheless, in the last years, and especially with
the advances in numerical simulations, a number of
 apparent inconsistencies
arose  between what is observed and what the CDM
model  predicts on small scales.
In this work, the current status of observations on galactic
and subgalactic scales is reviewed.
Furthermore, theory and observation are brought together in order to reveal
the nature of the inconsistencies  and,  consequently, to reveal their
severity. Lastly, the progress towards the resolution of each one
of these conflicts is briefly reviewed.
\end{abstract}

\maketitle
\tableofcontents

\section{INTRODUCTION}
There is something more in the Universe than luminous
matter. This additional,
{\it missing mass} component is  named {\it dark matter}. Deciphering
the nature of dark matter is among  the most important
cosmological problems seeking solution.
Since ordinary matter is baryonic, the
assumption that dark matter is baryonic  as well is the
obvious first assumption to make. Thus the diffuse, warm, intergalactic
medium, cold $H_{2}$,
Massive Compact Halo Objects 
(MACHOs~\footnote{ MACHOs  could be
stellar evolution remnants (white dwarfs, neutron stars, black holes) of
an early generation of massive stars ({\it population III} stars), or
smaller objects that never initiated nuclear burning, e.g., brown
dwarfs.}), etc., have been
proposed as baryonic dark matter candidates.
Nevertheless, not all of the dark matter can be baryonic,
due to constraints imposed on the contribution of baryons
to the critical density ($\Omega_{b}$) by Big Bang
Nucleosynthesis (Burles \emph{et~al.}, 2001a, 2001b),
as well as by the observed CMB anisotropies (Balbi \emph{et~al.}, 2000;
Lange  \emph{et~al.}, 2001;
Pryke \emph{et~al.}, 2002).
For example, assuming a baryon density 
large enough to account for the dark matter will lead
to CMB anisotropies that are larger than that observed.
Thus, at least part of the dark matter
has to be non-baryonic.

\begin{figure}
\epsfig{file=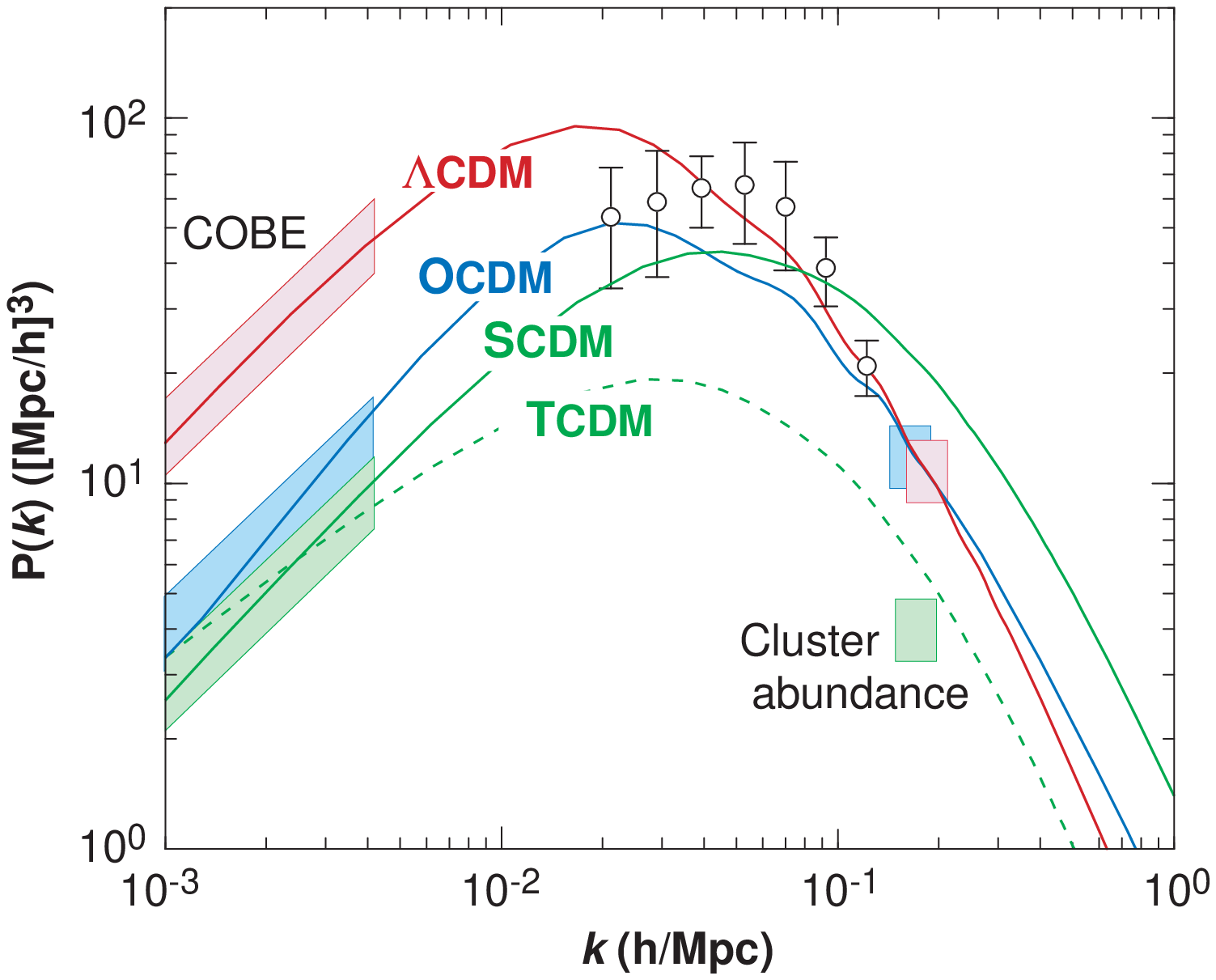, width=9.6cm, height=9.6cm}
\caption{\label{one} [Taken from Bahcall {\it et al.} (1999).] 
The power spectrum for several variants of the CDM model:
Standard CDM (SCDM), Tilted CDM
(TCDM), Open CDM (OCDM), and $\Lambda$CDM. The shaded areas on the left
represent COBE and CMB anisotropy measurements. The boxes on the
right are measurements of the cluster abundance at $z \simeq 0$. The data
points with open circles and 1$ \sigma$ error bars represent the APM
galaxy redshift survey.} 
\end{figure}

The first non-baryonic dark matter candidate studied was the
(light) neutrino,
a particle known to exist. Nevertheless, there is now significant evidence
against neutrinos as the bulk of the dark matter. More
specifically, neutrinos would be  hot dark matter.
Namely, they were relativistic when they decoupled. For
our purposes, what is even more important is the fact that they were
still highly relativistic when the horizon
encompassed for the first time the amount of matter
found in a typical galaxy
(Primack, 2001, and references therein).
It can be proven that
in a universe dominated by ordinary,
light neutrinos structure forms top-down,
that is,
large structures form first and smaller structures later by fragmentation
of the larger objects. This is opposite to what is inferred
from observation; for example, according to
CMB and cluster abundance measurements, the power spectrum is that of
the CDM model (see Fig.~\ref{one}), and thus, structure
has formed bottom-up, namely
small structures  formed first and then they merged towards the
creation of larger structures.

Thus, a lot of effort went to the so-called Cold Dark Matter
(CDM) (Blumenthal \emph{et~al.}, 1982; Peebles, 1982).
In the CDM model, dark matter consists of
non-baryonic, collisionless, cold particles. These particles are said
to be cold, since they  decoupled while 
non-relativistic. In addition, they
are assumed to  have had small primeval
(before structure formation) velocities, either because
they were thermally produced (exactly as in the case of hot dark matter),
but with a considerably high rest mass as, e.g.,
Weakly Interacting Massive Particles (WIMPS) are usually assumed to be,
or because they were (non-thermally)
created with momenta  well below
$k_{B}T$, e.g., axions
[for a  non-thermal WIMP production scenario,
see Lin \emph{et~al.} (2001)].
The most well-known WIMP, the neutralino, is predicted by the Minimal
Supersymmetric extension of the Standard Model (MSSM) of
particle physics (Haber and Kane, 1985). It is, by definition,
 the Lightest Supersymmetric
Particle (LSP) with a mass in the range of several tens of GeV up
to a TeV. Axions, on the other hand, are predicted by extensions
of the Standard Model that resolve the strong
CP problem (Turner, 1990).
They can occur in the early Universe in the form of a Bose
condensate that never comes  into thermal equilibrium. Axions
formed in this way are non-relativistic and can be a significant
dark matter contribution if their mass is $\simeq 10^{-5}$ eV. They
can also be produced from the decay of a network of axion strings
and domain walls [for more on axions,  see Kolb and Turner (1994)].

An essential part of a cosmological model  is
the  power spectrum of  density fluctuations that
it assumes (see Fig.~\ref{one}).
In the CDM model,  
structure grows out of random, Gaussian, adiabatic 
(curvature~\footnote{Curvature   fluctuations are fluctuations in the energy
density that can be characterized in a gauge-invariant manner as
fluctuations in the local value of the spatial curvature. The
name adiabatic comes from the fact that, in such perturbations,
 the corresponding
fluctuations in the local number density of any species with respect to
the entropy density   vanishes [see, e.g., Kolb and Turner (1994)].})   fluctuations.
These primordial density fluctuations are assumed to
 have the following power-law spectrum
\begin{equation}
P_{k} \propto k^{n}
\end{equation}
with $n$ being the spectral index. There are several constraints on $n$,
and the most widely  used value is $n=1$
that corresponds
to a  scale-invariant spectrum, known as the Harrison-Zel'Dovich spectrum
(Harrison, 1970; Zel'Dovich, 1972).

Numerous studies ( e.g., Davis \emph{et~al.}, 1985;
Bardeen \emph{et~al.}, 1986; Peebles, 1993, and references therein) have
been carried out regarding the evolved form of this
power spectrum. A simple form published,
valid 
after matter-radiation equality and before the
development of nonlinear structures,  is (Peebles, 1993, and
references therein)
\begin{equation}
P_{k} \propto \frac{k}{({1 + \alpha k + \beta k^{2}})^{2}}
\end{equation}
where
\begin{equation}
\alpha={8 / (\Omega h^{2}) {\rm Mpc}}, \hspace{1cm}
\beta= 4.7 / (\Omega h^{2})^{2} {\rm Mpc}^{2}.
\end{equation}
This form, even though not very detailed, reveals
the important feature of the spectrum: the spectrum bends
gently from $P_{k} \propto k^{-3}$ (large $k$, subgalactic scales) to
$P_{k} \propto k$ (small $k$, larger scales).

The model is definitely motivated by the inflation scenario, even though
it is not uniquely predicted by it. There are some
relatively simple versions  of inflation that predict exactly
the power spectrum that the CDM model has adopted. There are,
however, other versions of  inflation   that lead
either to quite
different forms for the spectrum,
or to isocurvature fluctuations
[see, e.g., Efstathiou and Bond (1986)].

Even so, cosmological models based on
the paradigm of an inflationary universe
with CDM, and either
a cosmological constant or quintessence (Caldwell \emph{et~al.}, 1998;
Carroll, 2001a, 2001b), have
recently enjoyed great success on large scales [for a review
on CDM successes, see  Bahcall \emph{et~al.} (1999), or
Wang \emph{et~al.} (2000)].
Impressively diverse data like
measurements of the CMB background radiation
(e.g., Lee \emph{et~al.}, 2001; Halverson \emph{et~al.}, 2002;
Netterfield \emph{et~al.}, 2002),
of the
abundances of deuterium and other light
elements (e.g., Burles \emph{et~al.}, 2001a), of the
absorption lines in the spectra of distant quasars (e.g., Efstathiou
\emph{et~al.}, 2000),
surveys in the positions
in redshift space of hundreds of thousands of galaxies (e.g., Peacock
\emph{et~al.}, 2001),
measurements of the brightness of distant supernovae (e.g.,
Riess \emph{et~al.}, 1998; Perlmutter \emph{et~al.}, 1999),
of the age of the Universe as measured from
the oldest stars  (e.g., Chaboyer \emph{et~al.}, 1998),
of the extragalactic distance scale as measured by distant
Cepheids (e.g., Madore \emph{et~al.}, 1998; Freedman \emph{et~al.}, 2001),
of the baryonic mass-fraction of galaxy clusters
(e.g., White \emph{et~al.}, 1993), of the present-day
abundance of massive galaxy clusters (e.g., Eke \emph{et~al.}, 1996;
Carlberg \emph{et~al.}, 1997; Bahcall
and Fan 1998),
of  the shape and amplitude
of galaxy clustering patterns (e.g., Wu \emph{et~al.}, 1999),
of the magnitude of large-scale coherent
motions of galaxy systems (e.g., Zaroubi \emph{et~al.}, 1997),  etc.,
all point to a single {\it new  cosmology} (Turner, 2002).
In this cosmology ($\Lambda$CDM), matter makes up at present less
than about one third ($\simeq 0.3$) of the critical density, and  a nonzero
($\simeq 0.7$)
cosmological constant ($\Lambda$) restores the flat geometry predicted by
most inflationary models of the early Universe. The present rate
of expansion is $H_{0} \simeq 70 $km s$^{-1}$ Mpc$^{-1}$,
baryons make
a very small fraction of the mass of the Universe ($\Omega_{b} \simeq
0.0195h^{-2}$), and the present day rms mass fluctuation, $\sigma_{8}$,
in spheres
of radius $8 h^{-1}$ Mpc  is of order unity.

Nevertheless,
an array of observations on galactic and subgalactic scales appears
to be in conflict with both analytical calculations and numerical
simulations  done  in the frame of the CDM model. These
conflicts are what this article is  about.
More specifically, I focus on the discrepancy that exists between
the theoretical predictions and the observations with respect to the
central mass distribution of galaxies, to the
shapes of galactic halos, to the existing substructure
in these galactic halos, and to the angular momentum of the
disks of  galaxies.
Throughout the paper, no distinction is made  between the standard
CDM and the $\Lambda$CDM models, unless their predictions differ
interestingly.

\section{MASS DISTRIBUTION IN THE INNER PARTS OF GALAXIES}
\label{massdist}
Cold dark matter halos are the result of a complicated
sequence of hierarchical mergers that
lead to a global structure set primarily by violent
relaxation.~\footnote{A collisionless relaxation process taking place
in time-varying
gravitational fields. For more details,
see, e.g.,  Binney and Tremaine (1994).}
Simulations of CDM halos (e.g., Navarro \emph{et~al.}, 1995, 1996;
Moore \emph{et~al.}, 1998, 1999a, 1999b;
Klypin \emph{et~al.}, 1999;
Ghigna \emph{et~al.}, 2000;
Jing and Suto, 2000)
result in halo density profiles,
for masses in the range $10^{7} M_{\odot} - 10^{15} M_{\odot}$,
that agree with a density profile  of the general
form introduced by Zhao (1996)
\begin{equation}
\label{profi}
\rho(r)=\frac{\rho_{o}}{\left(r /r_{o} \right)^{\gamma}
\left[ 1 + \left(r / r_{o} \right)^{\alpha}\right]^{({\beta - \gamma})
/ \alpha}}
\end{equation}
where $ \rho_{o}$ is the {\it characteristic density}
of the halo, and  $r_{o}$ is its {\it characteristic
 scale radius}. The characteristic density  is
defined as follows,
\begin{equation}
\rho_{o}= \rho_{crit} \delta_{c}
\end{equation}
where $\rho_{crit}$ is the critical
density today, and $\delta_{c}$ is the dimensionless overdensity
of the halo
that depends on the collapse redshift and on the mass of the halo.
The usual definition of $r_{o}$ is given in terms of the
logarithmic slope $d \log{\rho} /d \log{r}$ of the density profile. It
is defined as the distance from the
center of the halo where the logarithmic slope of the
profile equals $-(\beta + \gamma)/2$, with $\beta$ and
$\gamma$ the same as in Eq. (\ref{profi}).
Another quantity often used to characterize the dark matter halos
is the {\it concentration parameter c}. This parameter
is defined as the ratio of the virial radius, $r_{virial}$, of the
halo to its scale radius, $r_{o}$, namely $c = r_{virial} / r_{o}$.
The virial radius, on the other hand, is defined as the
distance from the center of the halo that roughly encloses
the region where matter is virialized.
Often, $r_{virial}$ is
taken to be equal to $r_{200}$, namely the radius of
a sphere of a mean interior density equal to $200 \rho_{crit}$.
Note that the characteristic overdensity $\delta_{c}$ and the concentration
index $c$ are not independent parameters; they are linked by the
requirement that the mean density within $r_{200}$  be equal to
$200 \rho_{crit}$.

Originally, it was found that the dark matter
density distribution in halos is well fitted
by the specific case of Eq. (\ref{profi}) known as  
the Navarro, Frenk and White
(NFW) (Navarro \emph{et~al.}, 1995, 1996, 1997) density profile
corresponding to
$(\alpha, \beta, \gamma)=(1,3,1)$.  Higher
resolution simulations that
followed
(Fukushige and Makino, 1997, 2001;
Moore \emph{et~al.}, 1998, 1999b;
Ghigna \emph{et~al.}, 2000;
Jing and Suto, 2000)
found somewhat
steeper profiles at small radii, with a
behavior in agreement with
the Moore \emph{et~al.} (1999b)
density
profile corresponding to $(\alpha, \beta, \gamma)=(1.5,3,1.5)$.
The two profiles differ essentially only in their
$r \rightarrow  0$ ($r < r_{o}$) behavior, with the Moore \emph{et~al.}
being steeper, whereas at larger radii ($r > r_{o}$) they
both behave roughly as $r^{-3}$.

Ongoing debate centers on the radial dependence
of the logarithmic slope, as well as on whether this slope indeed
converges to a well defined  central value. Several opinions
exist. For example, taking into account the conventional way of
defining the center of a halo, namely by finding the most
tightly bound particle, and given that the number
of particles per halo in the simulations
is finite, it becomes clear that, eventually, the choice
of the center of the halo may be somewhat arbitrary;
any departure from
the true center of the halo, which is very probable given
the currently feasible resolution,
will give the impression of a halo less
cuspy, namely less rapidly divergent at
$r \rightarrow 0$,
 than it actually is (Wandelt \emph{et~al.}, 2000).
 This is however only one simple
way of analyzing the problem. In general,
resolution and discreteness effects
can do either, namely they can lead to artificially low central
densities, or
to the formation of artificially dense
central cusps (Navarro, 2001).
To solve these problems, several
convergence studies (e.g., Klypin \emph{et~al.}, 2001;
Navarro, 2001; Power \emph{et~al.}, 2002) have been conducted.
These studies
consider the
effects of the numerical parameters of the simulations -- such as the
timestep, the number of particles, and the
gravitational softening~\footnote{Each {\it particle} in N-body
simulations stands for a large number of less massive particles.
For  the system  to be collisionless, one must ensure that
there will be no close encounters among the massive {\it particles}.
To prevent these collisions, one {\it softens} the force below a
certain distance, that is
sets the gravitational force equal to  a constant, instead of
$\propto r^{-2}$.} -- on the
density profile of the simulated halos.
As expected,   the two models were found to be similar 
at large radii (above $1 \%$ of the halo virial radius); in the
case of the NFW profile and  for small $r$ where the profile takes
the form $r^{-\alpha}$, it was found that the asymptotic slope
$\alpha=1$ is not obtained in the simulations
even at radii less than  $1\%-2\%$ of the
virial radius, whereas the Moore \emph{et~al.} profile was found to be a
very accurate fit for radii larger than $0.5 \%$ of the virial
radius (Klypin \emph{et~al.}, 2001).
Navarro (2001) found that
the radial dependence of the slope differs from that
 of the NFW profile,
that the NFW profile underestimates the density
at radii just inside $r_{o}$, and  that the Moore \emph{et~al.}
profile describes the simulated halos better than the NFW
profile in the
range $0.15 < r / r_{o} < 0.5$, even though it deviates
systematically at smaller radii.  Recently,
Power \emph{et~al.} (2002)
found that 
the logarithmic slope of the spherically-averaged density profile
is less than 1.2 and  that the
profile becomes increasingly shallow -- has smaller
and smaller logarithmic slope -- inwards, with little sign of
convergence to an asymptotic value in the inner regions.
From the above discussion,
if nothing else, it becomes clear that things are anything but
settled.

No matter what the exact value of $\alpha$ is, it is
a fact  that so far
CDM models
predict a  structure for  the halo density profiles that
behaves as $r^{-\alpha}$ at small radii, with $\alpha \geq 1$ ;
this profile diverges as $r \rightarrow 0$,  leading to   the
existence of a central cusp.
This is to be expected since CDM particles are by
definition moving slowly, and therefore there are no primordial
phase space
constraints~\footnote{This arises from an argument
given by Tremaine and Gunn  (1979). The basis for this
argument  is that the maximum coarse-grained phase space density
of halo dark matter cannot exceed the maximum primordial fine-grained
phase space density, which is conserved for collisionless matter.
Estimating the maximum phase space mass-density, one finds that
it only depends on the mass of the particle. In the case of a cold,
massive particle, this quantity is large, and thus in practice there is no
constraint.
}  that
could impose a cosmologically significant scale (Moore, 1994)
[as opposed, for example, to the case
of Warm Dark Matter particles (Alam \emph{et~al.}, 1990)].
This also implies that CDM is consistent  with
cusp - and not with core - dominated halos,~\footnote{Halos
with a central, essentially, constant density core.}
since the core radius will
be negligible (Moore 1994; van Albada, 1982).

This is what one expects from the CDM model
 regarding the central mass distribution
of halos. It is time to see what observation has to say about this topic.

\subsection{Rotation Curves}
\label{rotcurv}
As ironic as it is, the potential problems with the structure of CDM halos
were first highlighted by the  observations this theory was initially
designed to resolve: the flat rotation curves.  

\begin{figure}
\epsfig{file=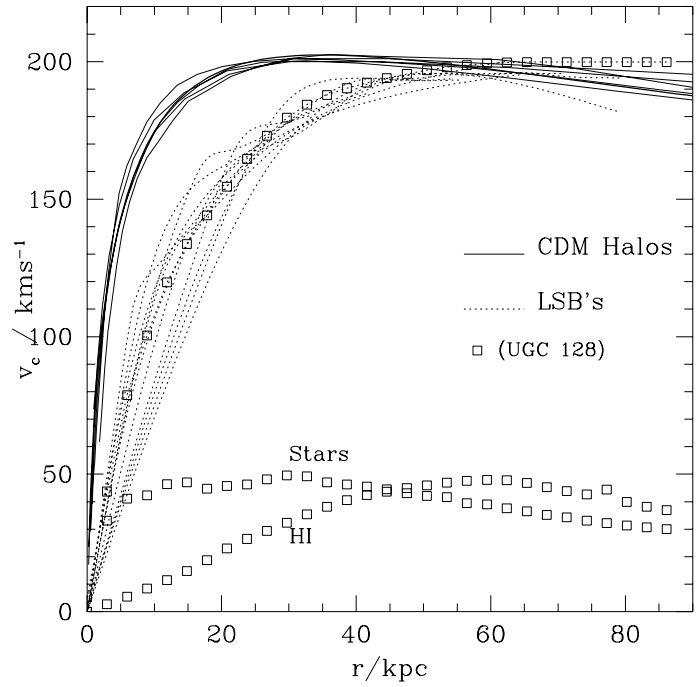, width=9.6cm, height=9.6cm}
\caption{\label{two} [Taken from Moore {\it et al.} (1999b).] 
Rotation curves 
of high resolution CDM halos (solid curves) and
LSB rotation curve data (dotted curves). The total rotational velocity
and the baryonic contribution from the stars and gas for a typical
LSB galaxy (UGC 128) are shown by open squares.
Note the steeper rise
of the model galaxy rotation curves compared with the observations.}
\end{figure}

More specifically,
the analysis of rotation curves
of some nearby dwarf galaxies~\footnote{Dwarf (elliptical or spiral) galaxies are
galaxies with considerably fewer stars and smaller sizes than their
normal counterparts. For instance,  Leo II is
a dwarf elliptical galaxy with approximately one million
stars, which means that its luminosity is about
equal to the luminosity of the brightest individual stars,
and a diameter of about 1.5 kpc. Note that the majority of
elliptical galaxies are dwarf galaxies (Morrison \emph{et~al.}, 1995).}
has indicated that
the steep rotation curves implied by CDM are hard to reconcile with the
observed shallow rotation curves.~\footnote{Steep and
shallow rotation curves mean
rapidly rising and  slowing rising rotation curves,
respectively (see Fig.~\ref{two}).}
Flores and Primack (1994)  found that  the rotation
curve data of the well-studied, gas rich dwarf spirals
DDO 154 and DDO 168  are inconsistent with
a density profile singular at the center. 
Burkert (1995) derived similar conclusions 
after  studying  seven  dwarf spiral
galaxies. Both studies determined
that these dark matter halos have a constant density core,
namely that they are core- and not cusp-dominated.
Another  study of nine low-luminosity, late-type galaxies,
in addition to  the
aforementioned, concluded that the constant density inner region (the core)
was comparable to, or even greater in size, than the
galaxy optical
extent (Borriello and Salucci, 2001; Salucci and Borriello, 2001). Recently,
Bolatto \emph{et~al.} (2002) have achieved  high
resolution combining  CO and H$\alpha$
observations in order to  construct the rotation curve of the
dwarf spiral NGC 4605; nevertheless, their results were similar
to the previous cases and thus did not support the
CDM model.

The so-called Low Surface Brightness galaxies (LSB)~\footnote{LSB galaxies
are conventionally defined as  galaxies that have
an extrapolated central disk surface brightness  in the blue band
fainter than 23 mag/arcsec$^{2}$ (Freeman, 1970). In practice, they
are a mixed group including objects as diverse as
giant gas-rich disks and dwarf spheroidals.}
present yet another
opportunity for studying
the dark matter distribution by means of rotation curves
-- even though with larger uncertainties in the observational
data compared  with dwarf
galaxies (Firmani \emph{et~al.}, 2000, and references therein).
These galaxies span a range of masses and
have a large fractional amount of dark matter inside their optical
region (Persic \emph{et~al.}, 1996; de Blok and McGaugh, 1997).
Thus, the contribution of the luminous matter
-- that is of the disk, bulge, and bar (if present) -- to the
gravitational potential is small.
This, along with the fact that their stellar distribution is almost a
perfect exponential,
simplifies considerably their
mass  modeling (Borriello and Salucci, 2001).

The analysis of 19 LSB HI rotation curves  by de Blok and McGaugh (1997)
indicated that they
rise less steeply  than their High Surface Brightness
(HSB)~\footnote{Galaxies with central disk surface brightness
$\mu_{B}\simeq 21.65\pm 0.30$ mag/arcsec$^{2}$[{\it Freeman's law},
(Freeman, 1970)].
Note that the distribution in surface brightness from HSB to LSB galaxies is
continuous, and the apparent
gap is only because of the convention used to define
the LSB galaxies.
}   counterparts of similar
luminosity, contrary to CDM predictions.
After this analysis, other
studies  (e.g.,  Moore \emph{et~al.}, 1999b; Firmani \emph{et~al.}, 2000)
came to verify that the observed rotation curves in LSB galaxies
rise less steeply than  predicted by CDM models (see Fig.~\ref{two}).
Furthermore, the behavior of the observed LSB rotation
curves requires a constant mass density in the
central region
(Moore \emph{et~al.}, 1999b; Firmani \emph{et~al.}, 2000;
de Blok, McGaugh, and Rubin, 2001; Moore, 2001), 
and this implies  core-dominated
halos; in fact, de Blok, McGaugh, and Rubin (2001) found that
the density distribution
profiles inferred from observations appear to be fitted accurately by
pseudo-isothermal profiles,~\footnote{Profiles that
exhibit the  singular isothermal behavior $ r^{-2} $  at large distances,
but have a constant density core, namely they are described by
the expression: $ \rho(r)=\rho_{o} \left[ 1+({r / r_{core}})^{2}
\right]^{-1}$, with $\rho_{o}$ and $r_{core}$ the core
density and radius, respectively.}
which  are core-dominated, whereas
the cusp-dominated CDM profile systematically deviates from the
data, often having a small statistical probability of being the
appropriate model. Recently, de Blok and Bosma (2002)
via their study of high resolution rotation curves
for 26 LSB and dwarf galaxies verified the aforementioned, finding
in addition that the
constant density cores are kpc-sized.

Moore \emph{et~al.} (1999b) reached similar
conclusions about a constant density central region
when they compared the scale-free shape of
the observed rotation curves
with simulation data.
For $\rho(r) \propto r^{-\alpha}$,
results from high quality measurements of rotation curves
indicate a best fit of $\alpha \simeq 0.5$, with significant scatter, which
is less divergent than the CDM profiles (Wandelt \emph{et~al.}, 2000,
and  references therein). For a large sample of LSBs
it  was found that the distribution of inner slopes
is peaked around $\alpha=0.2$, whereas $\alpha \simeq 1$
was found to be the case
only for the worst resolved cases (de Blok \emph{et~al.}, 2001). 
 Furthermore,  only extremely well-measured
rotation curves constrain simultaneously both
the slope $\alpha$ and the
concentration parameter (van den Bosch \emph{et~al.}, 2000).
Thus, in practice, the majority of observations
nowadays  do not resolve these two
parameters independently.
This means that we  can rephrase the problem in terms
of the concentration parameter: CDM models yield a wide range of
possible concentration
indices [with values 10-20 (Navarro
\emph{et~al.}, 1996; Bullock \emph{et~al.}, 2001b)],
but most
fits to rotation curves yield values well below (with values 6-8)
the predicted range in all types
of galaxies, even in the case
of the Milky Way (Navarro and Steinmetz, 2000a).

Another related issue is the value of the central density and 
the way it  varies from dwarf galaxies to clusters of galaxies. 
In fact, an analysis of data for dwarf and LSB
galaxies, as well as some clusters of galaxies, found that the halo
central density is nearly independent of the mass from galactic to galaxy
cluster scales with an
average of 0.02 M$_{\odot}$ pc$^{-3}$ (Firmani \emph{et~al.}, 2000),
whereas in the frame of CDM the central density is roughly proportional to
the mean cosmic density at the time of collapse; this means that small
mass galaxies, which collapsed sooner, when the cosmic density was high, 
are expected to have much higher densities than, e.g., clusters which formed
much later, when the Universe was less dense. Thus, according to
Firmani \emph{et~al.} (2000),
there is an additional disrepancy, that of the
central density dependence on the mass of the object. However, 
Shapiro \emph{et al.} (2002)
showed that when more and better data are used, then CDM and observation are
generally in agreement. 

In addition, the comparison
of observations of LSB galaxies with CDM predictions can be often
dismissed because of the limited resolution of the observed $H$I
rotation curves.
 The
early LSB rotation curves were obtained using the Very Large Array
(VLA)
and the Westerbork Synthesis Radio
Telescope (WSRT).
The relatively
large beams of these instruments resulted
in rotation curves with only a limited
resolution, and the effect of this on the obtained rotation curves
is known  as {\it beam smearing} (de Blok and McGaugh, 1997;
de Blok, McGaugh, and Rubin, 2001).
Re-examining the rotation curves for a sample of
LSBs,  taking this time into account the beam smearing, van den Bosch
\emph{et~al.} (2000) concluded that the observations
are of such low resolution that they place little, or no 
constraints on the inner shapes of the dark matter profiles. The
only galaxy for which they had very well-resolved data, and which
consequently was their most reliable object with respect to constraining
the dark matter profile, was found
to have an inner slope exactly in agreement
with the NFW density profile. Similar conclusions
were reached for dwarf galaxies.   
More specifically, van den Bosch and Swaters (2001)
analyzed and fitted the rotation curves of 20 late-type dwarf galaxies
by mass models with different
cusp slopes, ranging from constant density cores to $r^{-2}$ cusps.
Due to the large uncertainties, no unique mass decompositions 
were found, and it was thus concluded that the rotation curves
cannot be used to discriminate between halos that
at their central parts have a constant density or
are cuspy.
Furthermore, Swaters \emph{et~al.} (2000)
verified that optical data (H$_{\alpha}$)
sometimes indicate a steeper rise
of the  LSB rotation curves compared
with what was inferred for the same galaxies
via HI observations. The general shapes
of the  rotation curves of these LSB galaxies
were found to be almost identical to the
shapes of the rotation curves of HSB
galaxies,~\footnote{Almost identical shapes in the sense
that the    same fraction of the maximum rotation velocity -- the
velocity of the flat part of the rotation curve -- is reached at
the same distance from the center of the galaxy,
with the distance 
measured in units of the optical disk scale-length of the
galaxy (which is, in principle, different for different galaxies).
} in agreement with  CDM
predictions. 

Regarding other types of galaxies, apart from dwarf and LSB ones,
the rotation curves of a large number of
normal late-type galaxies  are consistent
with a fixed initial halo shape, characterized by a significant 
core inner region (Hernandez and
Gilmore, 1998) with  core radii much larger than a disk
scale-length (Salucci, 2001). Work
on barred galaxies (
Debattista and Sellwood, 1998, 2000; Weiner \emph{et~al.}, 2000)
has shown that any dark matter halo makes a negligible contribution to the
inner rotation curve,
 even after the formation  of the disk and the bulge.
Furthermore, there is empirical evidence against a systematic
difference between barred and unbarred galaxies, at least in the
case where these galaxies are HSB
ones (e.g., Mathewson and Ford, 1996; Debattista and Sellwood, 1998, 2000,
and references therein; Weiner \emph{et~al.}, 2000, and references therein).
They have similar  overall
HI properties, they appear to lie on the same
Tully-Fisher relation and to have similar
total mass-to-light ratios. These similarities,
in addition to that  halo dark matter
does not appear to be
the controlling factor with respect to whether a galaxy is barred or
unbarred,  imply that the dark matter distribution
in HSB barred and unbarred galaxies is similar, and thus
the conclusions derived regarding the dominance of the
luminous matter in the central parts of barred galaxies may well
be applicable in the case of their unbarred counterparts.
A similar generalization is sometimes done starting from what
has been concluded for  LSB galaxies and apply it to HSB galaxies.
The basis for such a generalization is
the fact that LSB galaxies  are  abundant,
span a wide range of masses,
show the same diversity as HSB galaxies (from giant disks
to dwarf spheroidals),
 have normal baryon fractions
(at least as many as HSB galaxies), lie on the same Tully-Fisher relation,
have similar structure with HSB galaxies -- namely
their light distribution falls of
exponentially -- and finally, they have properties that connect smoothly to
the properties of HSB galaxies. That they have  lower surface
brightness may result from larger than average angular momentum in the dark
matter component. On the basis that angular momentum is
uncorrelated with both
the environment and the halo structure, 
a generalization from  the LSB to the HSB galaxies can be valid
(see, e.g., Moore \emph{et~al.}, 1999b).

In fact, results similar to those obtained for LSB galaxies
were obtained for a large number of bright spiral galaxies, namely, the inner
part of these galaxies was found to  be dominated by luminous matter.
The impressive match  between  the inner rotation curves and the predictions
from luminous matter alone  often suggests
the model of the {\it maximum disk},
at least
in the case of the large spirals (Kent, 1986;
Palunas and Williams, 2000). In this model,
the rotation curve of the stellar  component is scaled to the
maximum value allowed by the observational rotation
curve, with the requirement  that the  dark matter density
be positive at all radii, in order to  avoid a so-called
{\it hollow halo}.
The fact that large spirals were found to be described adequately by
the {\it maximum disk} model indicates that
the light distribution
is an excellent predictor of the shape of the inner part of the rotation
curve, namely that dark matter might even be virtually
absent from the central parts of large spiral  galaxies.

Another possible problem regarding the rotation curves is
the {\it disk-halo conspiracy} (Bahcall and Casertano, 1985);
it refers to  that
if, according to the aforementioned, the luminous matter is dominant
in  the
central regions of galaxies, one  expects a feature in the galaxy
rotation
curves signifying the passage from luminous to dark matter as the dominant
source of gravity. Many galaxies are known nowadays, in
which the rotation curve drops  somewhat at the edge
of the visible disk,
but very rarely the drop
exceeds about $10 \%$ [see, e.g., Casertano and van Gorkom, (1991)].
Given that the observed feature
is very weak,  it is accurate enough to say that
observation verifies that the orbital circular
speed from the luminous matter at the center is  similar to the orbital
circular speed from dark matter at larger radii.
A conspiracy comes into existence on the basis
that  unless the dark and luminous
matter are related, the production of a flat rotation curve
requires that the initial conditions for both components, dark and
luminous, be finely tuned.
However, before concluding 
with respect to the existence of such a conspiracy,
one must take into account the compression of the dark matter by the
baryons during baryonic infall.  This process  provides a coupling
between the baryons and the dark matter and thus
  might result in
a total density distribution without any particular features or wiggles
to signify the transition from the one component to the other. Actually,
this was what Klypin, Zhao, and Somerville  (2001) found  when
they applied the standard
theory of disk formation within $\Lambda$CDM cuspy halo models,
including the effects of adiabatic compression.~\footnote{In the
simple case of a spherical system of particles on circular orbits,
adiabatic compression means compression conserving both the mass and
the angular momentum.}
In their models such a conspiracy is natural and
anything but surprising.

Summarizing, for several types of galaxies -- with LSB galaxies
the most numerous -- it appears
that their rotation curves
constitute observational
evidence for a smaller central mass concentration than
predicted by CDM. This is inferred by the fact that the central
parts of the galaxy rotation curves rise
less steeply than predicted.  These pieces of evidence would prove
the existence of a problem if, for
example, a unique mass decomposition could be
obtained from a rotation curve, or
if  observations were reliable, neither of
which is necessarily true. Lastly, the ``obvious" existence
of a disk-halo conspiracy,  might well be the result of incomplete
understanding and treatment of the physical processes involved.

\subsection{The Tully-Fisher relation }
\label{tfrb}
The Tully-Fisher relation (TFR) (Tully and Fisher, 1977) predicts
that the luminosity, L, of
a spiral galaxy  correlates with its  rotation velocity, $v_{rot}$.
This correlation is, essentially, a correlation
between the mass of the luminous galactic components with the rotation
velocity. Ideas regarding its origin may be grouped in two broad categories:
the one that sees the TFR as a result of self-regulated star formation in
disks
of different mass (e.g., Silk, 1997),
 and the other that sees the TFR as a direct
consequence of the cosmological equivalence between mass and rotation
velocity (e.g., Mo \emph{et~al.}, 1998).
The second idea, which is more related
to our approach, is
based on the existence of the virial radius,
which in turn, relates to the finite
age of the Universe.
More specifically, on dimensional grounds $v_{virial}^{2} \propto
G {M_{virial}/r_{virial}}$, and since in the
CDM context $r_{virial}
 \propto M_{virial}^{1 / 3}$ (Navarro  \emph{et~al.}, 1997),
 then $M_{virial} \propto v
 _{virial}^{3}$ where $v_{virial}$ and $M_{virial}$
are the halo circular velocity and  mass, respectively.
Under the assumption that
the disk rotation velocity, $v_{rot}$, is proportional
to $v_{virial}$, and that the total  stellar mass is  proportional
to $M_{virial}$, the total luminosity of the
galaxy will scale approximately as $v_{rot}^{3}$
[also see Mo \emph{et~al.} (1998), and Dalcanton \emph{et~al.} (1997)].

\begin{figure}
\epsfig{file=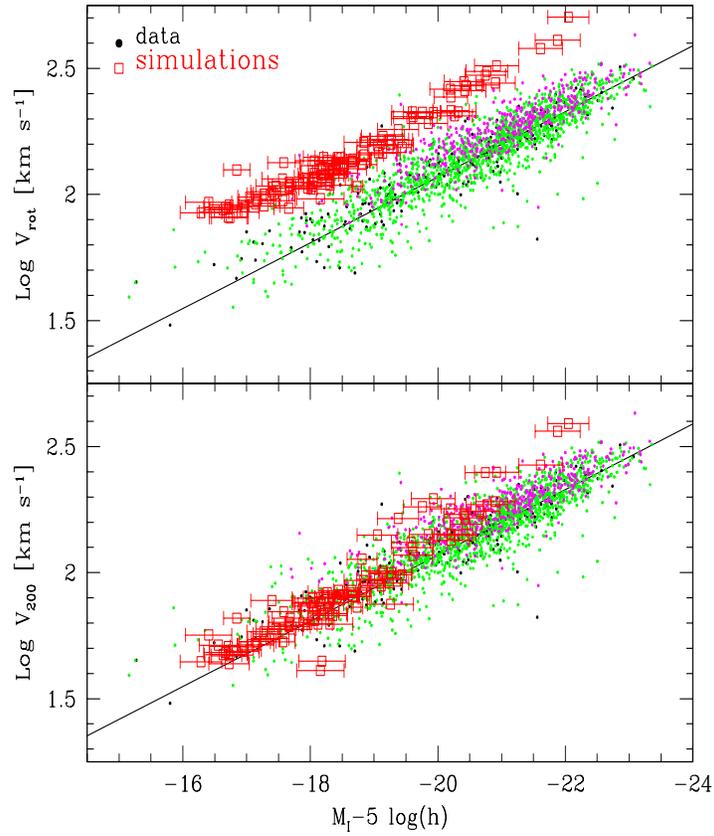, width=9.6cm, height=12cm}
\caption{\label{three} [Taken from Steinmetz and Navarro (1999).] 
I-band TFR at $z=0$.
Upper panel: simulations reproduce
approximately the slope and dispersion,  but not the zero-point of
the observational TFR. Lower panel: when using $V_{200}$ (halo
rotation velocity) all features of the observational TFR, even the
zero-point, are
well reproduced.}
\end{figure}

The expression for the observational TFR is of the form
\begin{equation}
L =A v_{rot}^{\beta}
\end{equation}
where $\beta$ is the slope and $A$ is
the {\it zero-point}.~\footnote{Note that
the TFR is plotted usually in a $\log{v_{rot}}$-absolute magnitude diagram,
hence the
{\it zero-point} denotes the intersection with the $\log{v_{rot}}$ axis
(see Fig.~\ref{three}).}
The observed values of $\beta$ range from about $2.5$ to about $4$,
whereas both $\beta$ and $A$ may depend
on the waveband (Mo and Mao, 2000, and
references therein). The TFR 
systematically steepens from the blue to the
red passbands and is surprisingly tight,
especially at longer wavelengths (e.g., Kauffmann \emph{et~al.}, 1993;
Sprayberry \emph{et~al.}, 1995;
Willick \emph{et~al.}, 1997).
For a fixed $v_{rot}$ the dispersion in absolute magnitude is less than
1 magnitude. It is believed that
observational errors and intrinsic
dispersion are, approximately, equally contributing
to this dispersion (Willick \emph{et~al.}, 1997; Sakai \emph{et~al.},
2000; Verheijen, 2001).
The choice of the location where $v_{rot}$ is measured is very important.
The best measure of the (optical) TFR $v_{rot}$, on the basis of
the repeatability of its measurement, the minimization of the TFR intrinsic
scatter, and the match with the radio (HI) linewidths (which
is important given that HI linewidths have defined the standard for most
TFR calibrations to date)  is the rotation velocity measured
at $2.2$ times the exponential scale length
of the galactic disk. Note also that this is 
where the contribution of
a pure exponential disk
to the circular velocity attains its maximum, and it is at that radius
that TFR velocities are typically measured (Courteau, 1997).

With respect to the TFR
there are three major observational features  a model must
reproduce so that it  be considered successful: the slope, the
small dispersion, and the zero-point.
CDM numerical simulations
(Mo \emph{et~al.}, 1998;
Steinmetz and Navarro, 1999;
Mo and Mao, 2000;
Navarro and Steinmetz, 2000a, 2000b)
manage to reproduce a slope
in fairly good agreement with observational
data in many passbands.
The dispersion of the numerical TFR is
 about half of that observed (see Fig.~\ref{three}-upper panel).
Given that numerical calculations are free of observational dispersion,
and taking into account the aforementioned regarding the contributions
to the dispersion in the observational TFR, it seems
that the model  reproduces well the observed dispersion.
The problem is the zero-point. The numerical studies
result in one and the same conclusion: the zero point is offset by 1 to 2
magnitudes, depending on the cosmological parameters.
More specifically,
for a specific $v_{rot}$ the model galaxies in the numerical
simulations appear to be 1 to 2 magnitudes
fainter than the observed galaxies
(Mo \emph{et~al.}, 1998;
Steinmetz and Navarro, 1999;
Mo and Mao, 2000;
Navarro and Steinmetz, 2000a, 2000b).
This has also been hinted at by some
semianalytic CDM models of
galaxy formation (e.g., Kauffmann \emph{et~al.}, 1993;
Cole \emph{et~al.}, 1994).
In other words,
the numerically obtained $v_{rot}$
at a given luminosity is about $40 \%$ to $60 \%$ higher than
observed (Steinmetz and Navarro, 1999).
This result seems to be independent of the disk mass-to-light
ratio, and even with the extreme assumption that
all baryons in a dark
halo are turned into stars, the resulting disks are still
 about 2 magnitudes fainter
 than observed,  for a given $v_{rot}$ (Navarro and Steinmetz, 2000a).
This, along with the fact that $70\%-90\%$ of all baryons inside the
 virial radius are confined within the luminous extend of
 the disk, at zero redshift (Steinmetz and Navarro, 1999),
mean that the model galaxies  are already almost as bright as they can be;
thus, making the model galaxies even brighter to match the observations
is not an option.
Note though 
that the correct zero point can be reproduced when $v_{virial}$
is used
instead of $v_{rot}$
(e.g., Steinmetz and Navarro, 1999; also see Fig.~\ref{three}-lower panel).
Apparently, according to the aforementioned, $v_{virial}$ is
$40 \%$ to $60 \%$ lower than $v_{rot}$.

The large difference  between the two velocities is
expected in CDM models. The mass of the central galaxy in a halo-galaxy
system depends  on the efficiency of gas cooling which, in turn, depends
on the mass of the halo; the less massive the halo, the more efficient
in assembling baryons and dark matter, which is drawn by
the baryons, into galaxies.~\footnote{This is true according to
expectations from theoretical models  where the mass of the
central galaxy is determined by
the efficiency of gas cooling (Navarro and Steinmetz, 2000b).
}
Systems that collect a
large fraction of the available baryons into
a central galaxy have their rotation speeds increased substantially over
and above the circular velocity of their surrounding
halo  (Mo \emph{et~al.}, 1998;
Steinmetz and Navarro, 1999;
Mo and Mao, 2000;
Navarro and Steinmetz, 2000a, 2000b).
Given that  $v_{virial}$ depends on the total halo mass
and that $v_{rot}$ depends on
the total mass, both luminous and dark,  that lies inside the radius
where $v_{rot}$ is measured, it seems inevitable
that in order to lower $v_{rot}$ so that it approaches
$v_{virial}$ and thus reproduces the correct TFR
zero-point, one must reduce the mass of the halo concentrated in
the central regions. Namely, the zero-point discrepancy
translates into 
a problem with the large central concentrations of the dark matter halos
predicted in the CDM context.
For the numerically derived  TFR
to coincide with the observational one, the dark matter in the innermost
few kpc of  galaxies must be reduced
by a factor of 2 to 3 (Navarro and Steinmetz, 2000b)
in terms
of the concentration parameter, a reduction by a factor of $3 $ to
$5$ (Navarro 1998; Navarro and Steinmetz, 2000b)
compared
with the CDM values [$c =15-20$ (Navarro \emph{et~al.}, 1996)],
and a value of $\simeq 3$, referring
to the NFW profile
(Mo and Mao, 2000), have been reported.

It is noteworthy, however, that there are many points
that need to  be clarified.
For
example, Eke \emph{et~al.} (2001)  reported
that the results regarding the TFR zero-point
presented
by Navarro \emph{et~al.}  (2000a)
are wrong due to the high
power spectrum normalization used.
A higher normalization ($\sigma_{8} >1.14$) 
means a higher collapse time, and consequently, a higher density
and concentration which can be 
the source of the zero-point discrepancy. Indeed, Eke \emph{et~al.} using
$\sigma_{8}$=1.14, managed to reduce the offset to about 0.5 mag with
respect to the observational TFR.
They attributed that remaining offset
to the fact that in their simulations star
formation occurred sooner than it should -- this was hinted by
the fact that the simulated galaxies were slightly more red than
their observed TFR counterparts -- and  claimed
that correcting for
this effect can  eliminate the offset entirely.
This is an example of how crucial including a
realistic modeling  can be.

Another  aspect of this topic is the redshift evolution of the
TFR. Sample selection and details of the
observational technique used make the situation  unclear.
Claims that the TFR either brightens or dims at modest redshifts
($z \leq 1$) can be found in
literature  (Vogt \emph{et~al.}, 1996, 1997; 
Rix \emph{et~al.}, 1997;
Hudson \emph{et~al.}, 1998;
Simard and Pritchet, 1998).
Furthermore, it appears that whether the TFR dims or brightens depends
upon the parameters used. For example, Vogt \emph{et~al.} (1996, 1997)
reported that the TFR barely brightens only for the low
value of $q=0.05$, whereas the
same data for $q=0.5$ are consistent with a slight dimming, with
$q$ being the deceleration parameter.
The CDM model TFR  brightens
at $z=1$ by $\simeq 0.7$ or $0.2$ mag in the B-band,
depending on whether $v_{rot}$ or
$v_{virial}$ is used, respectively. Whether this is
inconsistent
with observation or not, depends again on the parameter values
and observational data one uses. Adopting
$q=0.5$ and the data from Vogt \emph{et~al.} (1996), there
appears to exist an inconsistency,
which,
nevertheless,  has been attributed so far to the specific star formation
algorithm used (e.g., Steinmetz and Navarro, 1999). In addition,
it appears that the model TFR follows the
observed dimming at passbands that are insensitive to star formation, e.g.,
the K-band (Steinmetz and Navarro, 1999).
At present, there are many things that need to be understood
with respect to
the redshift evolution of the TFR. In the future, however,
the TFR redshift evolution will be  a powerful
tool that can be used to test the CDM model.

Last, but not least, with respect to the TFR, one should mention what in
literature appears as the
{\it surface brightness conspiracy} (e.g., Evans, 2001),
causing
problems not only in the CDM model, but also in the very existence of dark
matter
in general.
The observational fact  is  that  the
rotation velocity, $v_{rot}$, is similar
for all  galaxies of a given luminosity, no matter
how widely spread the luminous material is. Namely, LSB galaxies
lie on the same TFR as HSB galaxies (e.g., Sprayberry \emph{et~al.}, 1995;
Zwaan \emph{et~al.}, 1995),
even though with somewhat
greater scatter (e.g., Sprayberry \emph{et~al.}, 1995;
McGaugh \emph{et~al.}, 2000).
On dimensional grounds
$v_{rot}^{2} \propto G{M / D}$, where $M$ is the mass (both luminous
and dark) of the galaxy and  $D$ is its characteristic size.
For two galaxies, one HSB and one LSB,
with the same luminosity L,
and with characteristic sizes $D_{HSB}$ and
$D_{LSB}$,~\footnote{Note that it has to be $D_{HSB} < D_{LSB}$
since on dimensional grounds $S \propto {L /  D^{2}}$, with $S$
and $L$ 
denoting the surface brightness and luminosity, respectively,
and given that $S_{HSB} > S_{LSB}$, by definition.
} respectively, to have
the same $v_{rot}$, the overall $M/L$ must increase with decreasing surface
brightness  in just the right way~\footnote{In a way so that the
increment of $D_{LSB}$ be compensated by an increment in $M$.}
so that the tight $v_{rot}-L$
correlation observed be justified.   
This can be done by assuming either that the $M/L$ of the stellar population
varies with surface brightness, which  seems unlikely to be the
case (de Blok and McGaugh, 1997), or
that the dark matter fraction increases
with decreasing surface brightness.
This in not a problem in the case of the LSB galaxies,
given that they are dominated by dark matter. 
The potential problem has to do  with
the bright galaxies whose inner
parts are dominated by the luminous matter; eliminating the surface
brightness dependance in the case of these bright galaxies requires
some fine tuning.

In conclusion, two out of the three  features of the observational
TFR are easily reproduced in the frame of CDM.  The zero-point
appears to be more difficult to predict correctly. For a specific
galaxy luminosity, the CDM predicted rotation velocity is  higher
than observed; this can be attributed to the high
central mass concentration predicted by CDM. The lack of
realistic modeling and other factors, such as inconsistencies
in the calculations, may be held, at least partially, responsible for
the zero-point discrepancy.

\subsection{Barred galaxies}
Bars are seen in optical images of roughly  $30 \%$ of
all disk galaxies (Sellwood and Wilkinson, 1993),
and the fraction of strongly barred
galaxies rises to over $50 \%$ in the
near-IR (Eskridge \emph{et~al.}, 2000). These
numerous objects constitute a useful
laboratory where the
dynamics of the central regions of galaxies may  be
probed and, thus, where the CDM hypothesis can be tested.

A way to  test CDM using
 this class of objects concerns  the  pattern speeds
of the rotating bars.
The usual way to characterize the rotation rate of a bar
 is the ratio $R= {d/\alpha}$,
where $d$ is the corotation radius and $\alpha$ the
bar semimajor axis. More precisely, $d$ is the distance from
the center to the Lagrange point on the bar major axis where the
gravitational attraction balances the centrifugal acceleration (in the frame
rotating with the bar). Theoretical arguments on the basis of bar stability
require $R >1$ (Contopoulos, 1980),
whereas  that $R$ can also equal unity
has been reported [for a review, see Sellwood and Wilkinson (1993)].
A bar can be classified as fast or slowly rotating based on its
R value.
Fast rotating bars  are the ones with low
values
of $R$.
For example, a classification
of bars with respect to this criterion treats bars with $1 \leq R
\leq 1.4$ as rapidly
rotating  and bars with $R>1.4$ as slowly
rotating ones (Debattista and Sellwood, 2000).
More generally, characterizing a bar as slow  means that it has an $R$
value substantially greater than 1.

Both numerical and analytical arguments
exist that lead to the conclusion 
 that if there is substantial dark matter in the bar region, the
rotation  of the bar is slowed down on a time scale of a few
rotation periods (e.g., Weinberg, 1985; Debattista and Sellwood, 1998, 2000;
Athanassoula and Misiriotis, 2002).
Along
the same lines with these arguments,
this is to be expected since CDM,
being dissipationless, rotates much more slowly than the
galaxy  baryons and does not participate fully in the galactic bar. The
response of the dark matter to the bar is out of phase with it, and this
provides a drag force that is usually referred to as dynamical friction.
This dynamical friction
is at its most severe when both dark matter and baryons
provide comparable mass (Binney and Evans, 2001).  This is, more
or less, the conventional
picture with respect to how the rotation of a bar is expected to
change with time due to its interaction with  the surrounding matter.
This picture is true, as long as the bar is a solid rotator.
In this case, a bar that loses angular momentum will inevitably
slow down. However, as pointed out by
Valenzuela and Klypin  (2002),
real bars may behave differently because they are not solid
rotators, but they consist of particles; when the bar loses
angular momentum it is the individual particles that suffer this loss
and  that move to smaller radius trajectories  that,
nevertheless, correspond  to higher
angular frequencies. The  bar speed is expected to be proportional
to the angular frequency of the particles, and since  losses of angular
momentum lead to higher orbital frequencies, it appears possible that
a real bar can speed up  while losing angular momentum, contrary to what
is usually assumed on the basis of  a solid bar rotator.
In the same study, Valenzuela and Klypin
also  show  that the results obtained
 via previous numerical simulations, which resulted in
  bars that slowed down fast (even though
far slower  than analytically predicted),  may not be accurate
due to the  initial setup used and 
the numerical resolution limitations.
More specifically,  in the low resolution simulations
they
 find that  bars appear to form from the very beginning -- long
before they appear  at higher resolution simulations -- and that
these bars
slow down relatively fast, at least compared with what happens in
the higher resolution cases where the pattern speed of the bars
appears to remain unchanged over billions of years. With respect to the
initial setup, they point out the fact that the numerical simulations
that the conventional picture for bar pattern speed evolution
is based on, do not use realistic halos, neither with respect to the
halo
virial extent -- usually halos in these simulations
are  truncated  at radii well below the virial
radius -- nor with respect to the halo  density profiles, since the
profiles used were not the ones suggested by CDM.
In terms of numbers, the disagreement is expressed as a
2.0 to 2.6 range for the $R$-values based on conclusions
derived, e.g.,  by Debattista and
Sellwood (1998, 2000), and
a 1.2 to 1.7 range based on the study carried out by
 Valenzuela and Klypin  (2002) which is, as we will see, partially
 in agreement with observations.

The number of galaxies with measured $R$ is  quite small. The main
difficulty is that it requires knowledge
of the bar angular speed, $\Omega$, which is hard to determine
from observational data, even though
reliable methods have been proposed and
been used (e.g., Tremaine and Weinberg, 1984; Dehnen, 1999).
Among the most reliable
results reported so far are the ones
obtained using  the Tremaine-Weinberg (1984)
method.
The method has yielded the  values
$R=1.4 \pm 0.3$~\footnote{The results published
are $69 \pm 15$ arcsec for the
corotation distance and $50$ arcsec for the bar extend from which
I estimated $R$.
}
for NGC 936 (Merrifield and Kuijken, 1995), and
$ R=1.15^{+0.38}_{-0.23}$ for NGC 4596
(Gerssen \emph{et~al.}, 1999). Indirect,
but reliable  estimates have been accomplished by matching hydrodynamical
and stellar dynamical simulations with
observations (e.g., Lindblad and Kristen, 1996;  Lindblad
\emph{et~al.}, 1996;  Englmaier and Gerhard, 1999;  Weiner and
Sellwood, 1999; Hafner \emph{et~al.}, 2000).
The typical range of values for $R$ was found to be from
1.1 to 1.3, approximately.
Namely, the existing observations
unanimously indicate that $R$ is small, which means
  that the corotation
point lies typically just beyond the bar's optical edge, and thus the
bars are rotating fast. According to the conventional picture, bars can
maintain the  high pattern speeds observed
only if the disk provides most of the
central attraction in the inner regions ({\it maximum disk},
see Sec.~\ref{rotcurv}).
It has been reported that at most
a $10 \%$ contribution by spherically or
axially-symmetrically distributed dark matter can be allowed within
the bar region to avoid conflict with the observed kinematics
of certain barred galaxies
(Sellwood and Kosowsky, 2000, and references therein).
In the case of our Galaxy,  a barred galaxy,
even at the Solar circle the halo
contribution to the radial force has been calculated to be
only $\simeq 20 \%$ (Englmaier and Gerhard, 1999).
According to these results,
again, there is no room for highly centrally concentrated
dark matter halos.
Moreover, as before, the generalization from
HSB barred galaxies to their HSB  unbarred counterparts 
seems rather attractive. Empirical evidence  has been  presented
against a systematic difference with respect to the
dark matter fraction of these two types of HSB
galaxies (Mathewson and Ford, 1996; Debattista and Sellwood, 2000,
and references therein).
Thus,  the conclusion that the aforementioned  regarding
the dark matter component hold for
all bright galaxies,  and not only for barred ones, has been presented
by some authors, e.g., Debattista and Sellwood, 2000.

Note, however, that  the 1.2 to 1.7 range for $R$ found by Valenzuela and
Klypin (2002) is, as already mentioned, in partial agreement
with the above, measured $R$ values.
Furthermore, Klypin, Zhao, and Somerville (2001)
found  that $\Lambda$CDM
models of,  e.g.,  the Milky Way  are  able to
reproduce the  observed kinematics, while satisfying
numerous
constraints in the solar
neighborhood (e.g., surface density constraints). They also
find that none of their models appears to be
dark matter dominated in the central region, in accordance with
what one expects for a galaxy like the MW, and that the sustenance
of a rapidly rotating bar appears to be possible,
as long as some transfer of angular momentum from the baryons to the
dark matter is included.

\subsection{The microlensing optical depth to the galactic center}

A crucial difference between baryonic and particle dark matter is that the
former can cause microlensing events, whereas the latter
 cannot.~\footnote{Referring to the smooth
component of the dark matter which, by definition,
has no substructure and thus cannot cause microlensing.
}
This fact can
be used to derive quantitative conclusions about the baryonic mass alone.

Extremely high microlensing optical depths towards the
galactic bulge -- either exactly towards Baade's
window ($l=1^{\circ}$ and
$b=-4^{\circ}$), or towards directions very close to
it -- have been measured.
The optical depths
reported so far lie in the range $(1.4-4) \times 10^{-6}$
(e.g.,
Udalski \emph{et~al.}, 1994;
Alcock \emph{et~al.}, 1995, 1997, 2000;
Zhao \emph{et~al.}, 1995;
Klypin, Zhao, and Somerville, 2001, and references therein),
with the most recent analysis   to conclude that the
optical depth is in the range $(1.4-2) \times 10^{-6}$
(Klypin, Zhao, and Somerville, 2001, and references therein).
All these
observational estimates  are nearly an order of magnitude
higher than the values  theoretically anticipated, which lie
typically in the range $(0.1-4) \times 10^{-7}$
(e.g., Griest \emph{et~al.}, 1991;
Paczynski, 1991; Kiraga and Paczynski, 1994),
except from the study carried out  more recently
by Klypin, Zhao, and Somerville (2001)  that results in the
  conclusion that the microlensing
optical depths  for their
model galaxies  can be of the order of
$10^{-6}$.

The microlensing optical depth increases with both the surface density of
bulge stars  and the effective depth of the bulge along the line of
sight.  These dependencies can be used 
to derive the contribution
to the rotation curve by the stellar disk and the interstellar medium.
To find the contribution
of the relatively unexplored bar, one must assume a bar massive enough
that
the total baryonic matter in the inner galaxy reproduces the measured
optical
depth. The other contribution to the rotation curve will come from the
dark
matter component. A cuspy CDM halo can be constructed in the case of the
Galaxy,
that should be normalized
using the appropriate data, for example, the local surface density,
which can be derived from kinematics of stars in the solar neighborhood.
Thus, the contribution of the CDM component to the rotation velocity
can be found, and the total  rotation velocity, which is to be compared
with the observed one, can be obtained by adding in quadrature
the two contributions, the  luminous and the dark matter ones.

This procedure in several levels of completeness
has been followed, e.g., by
Binney \emph{et~al.} (2000) and Binney and Evans (2001) for the Galaxy.
A unanimous conclusion has been reached: almost
all the matter along the lines of sight to the galactic bulge must
be capable of  causing microlensing,  and thus
it cannot be particle dark matter. Binney and
Evans (2001)
conclude that even the least
concentrated CDM halo profile used ($ \alpha =0.3$) is ruled out
since it violates the constraint of the observed rotation curve,
especially the
inner part that appears to rise extremely steeply.
In their study, no halo with
a cusp as  steep as or
steeper than $r^{-0.3}$  is viable.
According to the same study, to obtain an optical depth of order
$10^{-6}$ using an NFW halo, one must push all the related parameters
to their very extremes, i.e., one must use the smallest possible dark matter
surface density consistent with the observed rotation curve.
But, as pointed out by Klypin, Zhao, and Somerville (2001),
these conclusions are
based on a quite non-realistic treatment of both the dark matter and the bar.
For example, Binney and Evans 
used the unmodified NFW profile, ignoring
thus the effects that adiabatic compression has in the inner density
profile. Furthermore, Klypin, Zhao, and Somerville 
have modeled their bars in far greater detail, so that numerous
observational constraints be satisfied, and have used more 
detailed treatments of microlensing events compared with Binney and
Evans. All these led the two studies to totally different
conclusions, Binney and Evans to the conclusion that cuspy CDM halos
are inconsistent with microlensing data, and Klypin, Zhao, and Somerville
to the conclusion that
such an inconsistency is nonexistent.

Summarizing with respect to the apparent central mass discrepancy, there
is a large number  of observations indicating
that the CDM model predicts a higher
central mass concentration than  observed. This translates into a steeper
than observed rise of the CDM predicted inner rotation curves,
a higher than
observed zero-point of the TFR, into bars that rotate more slowly
than observed, and  into microlensing optical depths towards the galactic
bulge an order of magnitude lower than measured. However, all these are
pieces of evidence that there is something wrong with CDM, as long as
1. observations are reliable and are  interpreted in the correct way,
2. the assumptions made to derive the CDM theoretical
predictions are correct
and consistent,  and  3. the simulations are reliable with respect to both
resolution and convergence, and with respect to their realistic and
complete treatment  of the related physical processes.
We saw, using specific examples, that none of these prerequisites
is necessarily true. Thus, not only there is no proof for a discrepancy,
but also the
strength of the existing evidence is questionable.

\section{HALO SHAPES}

\begin{figure}
\epsfig{file=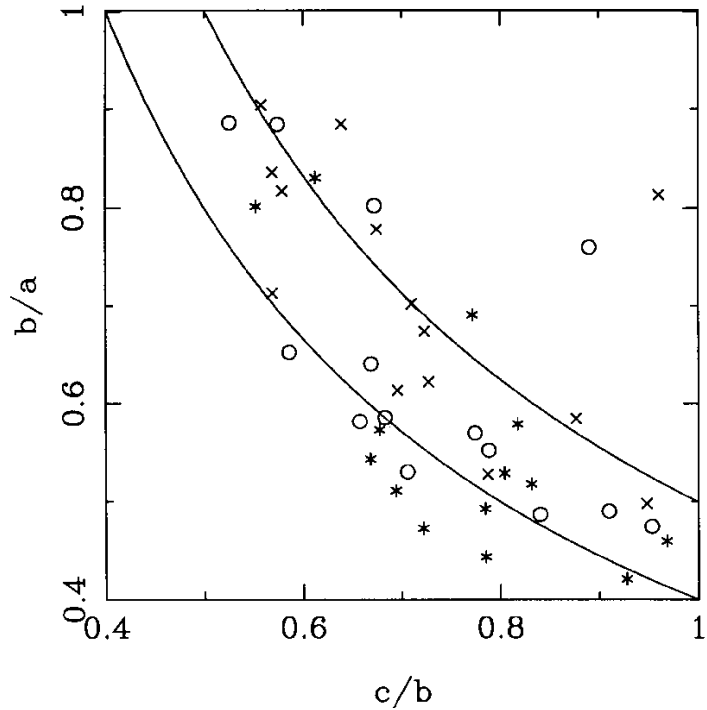, width=9.6cm, height=9.6cm}
\caption{\label{four} [Taken from Dubinski and Carlberg (1991).] 
Distribution of axial ratios for CDM halos (collisionless
simulations). Axial ratios within 25 kpc (asterisks), 50 kpc (circles),
and 100 kpc (crosses) are displayed. The solid curves represent
ellipsoids with $c / \alpha$ =0.4 and 0.5. In the inner regions
($<25$ kpc) the halos are very flat and prolate. The shapes measured at
larger radii represent oblate and prolate forms in approximately
equal numbers.}
\end{figure}

Dissipationless halo formation
leads to strongly triaxial halos, as revealed by numerical simulations
(e.g., Frenk \emph{et~al.}, 1988; White \emph{et~al.}, 1990;
Dubinski and Carlberg, 1991; Warren \emph{et~al.}, 1992;
Jing and Suto, 2002).  In fact, triaxial modeling
has been found to improve significantly  the fit to the
simulated profiles, at least for the relatively relaxed
halos, compared to the usual spherical model (Jing and Suto, 2002).
Both that there is a slight
preference for prolate
(e.g., Frenk \emph{et~al.}, 1988; Warren \emph{et~al.}, 1992) -- especially
in the inner regions -- configurations,
and that there are roughly equal numbers of halos
with oblate  and prolate forms~\footnote{Oblate: ${c/b}
< {b/a}$, prolate: ${c/b} >{b/a}$,
with $a$, $b$, $c$ the major, intermediate, and minor  axis of the triaxial
ellipsoid, respectively.}
have been reported (e.g., Dubinski and Carlberg, 1991).
These halos are highly
flattened. The flattening is usually expressed
using the ratio $c/a$, where $c$ and $a$ stand for
the minor and the major axis of the triaxial ellipsoid, respectively.
The typical value of this ratio, as calculated by numerical simulations,
is $\simeq 0.5$, leading to the above  conclusion that
CDM halos are highly flattened (Frenk \emph{et~al.}, 1988;
Dubinski and Carlberg, 1991; Warren \emph{et~al.}, 1992).~\footnote{The
${c/a} \simeq 0.5$ case also appears
in literature  as {\it 2:1 flattening}. It
is also worth noting that a halo of dark matter particles that is
flattened
more than about $3:1$ could not survive and would puff up
through
{\it dynamical bending instability} (Merritt and Sellwood, 1994).}
A mean value of the ratio ${b/a} \simeq 0.71$ has
been reported
(Dubinski and Carlberg, 1991; Dubinski, 1994), with $b$ denoting
the intermediate axis of the
halo (see also Fig.~\ref{four}).
Including the dissipative infall of gas -- which results
in the formation of the
luminous part of the galaxy -- causes the prolate halos to transform
from prolate triaxial to oblate triaxial, at least at the inner parts.
The 2:1 flattening is approximately preserved.  At their central
parts these halos are spheroids, as opposed to their outer parts where, at
least in a statistical sense, they become more triaxial
and twisted (Dubinski, 1994; Sellwood and Kosowsky, 2001).
Furthermore, in this case ${b/a} \geq 0.7$ (Dubinski, 1994).
The shapes of the triaxial halos formed in the CDM model
are supported by
anisotropic velocity dispersion rather than by angular momentum, since
their rotation alone is not sufficient to account for their flattening
(Warren \emph{et~al.}, 1992).
In the $\Lambda$CDM context, the ratio $c/a$ was found to be $\simeq
0.70$ to within $30h^{-1}$kpc with a scatter of about $\pm 0.17$ about the
mean, with a long tail skewed towards highly flattened
objects (Bullock, 2001b).

From the very first CDM  studies of halo shapes
it  became clear that there are some
similarities between halos and  elliptical galaxies. Elliptical
galaxies are also believed to be triaxial bodies supported by pressure
anisotropy (e.g., Binney, 1976; Frenk \emph{et~al.}, 1988).
Furthermore, it
is believed that
if the history of elliptical
galaxy formation is one of a hierarchy, involving lumps consisting
mostly of stars, then the dynamics of halo and elliptical
galaxy formation are  similar (e.g., Zurek \emph{et~al.}, 1988), 
 and their shape distribution
is expected to be the same
(e.g., van Albada, 1982; Aguilar and Merritt, 1990). Comparisons
though, between real elliptical  galaxies
and CDM halo shapes, concluded that  the former 
are much rounder (less elliptical) than the latter
(e.g., Frenk \emph{et~al.}, 1988; White \emph{et~al.}, 1990; Dubinski
and Carlberg, 1991).
Warren \emph{et~al.} (1992)  argued  that this
difference between the shape distributions of elliptical
galaxies  and halos
is expected.  Their argument is  based on
the dynamical friction that dense stellar systems undergo when
moving through the dark matter during a merger, and
which  results in a one-way transport of angular momentum and kinetic
energy from the orbits of the dense stellar systems to
the less dense dark matter
halo.

Even though the number of measurements is
very small,
mainly due to the lack of visible tracers that can probe the gravitational
potential around galaxies,
the predictions of the CDM model regarding the halo shapes, such
as the 2:1 flattening, 
are nowadays generally thought to
be consistent with
constraints on halo shapes inferred from observation
(Sackett, 1999; Merrifield, 2001; Hoekstra \emph{et~al.}, 2002).
Nonetheless,
there  are some exceptions. The halo of NGC 2403 seems to become more
nearly
axisymmetric at large radii
(Sellwood and Kosowsky, 2001, and references therein).
Sackett \emph{et~al.} (1994) found for the polar ring galaxy
NGC 4650A that the inner dark matter halo is as flat
as $0.3 \le {c / a} \le 0.4$. Olling (1996), on
the assumption of an
isotropic gas velocity dispersion tensor,
estimated that the inner
halo of NGC 4244 is as flat as ${c / a} \simeq 0.2^{+0.3}_{-0.1}$.
The halo of NGC 3198 was found to be closely
axisymmetric at all radii outside
the disk (Sellwood and Kosowsky, 2001, and references therein). The
gas ring
surrounding the early-type galaxy IC2006 was found to
have an ellipticity equal to $0.012 \pm
0.026$, namely it was found to be essentially circular,
whereas the departure from
axisymmetry of the potential was estimated
to be $\le 1 \%$ (Franx \emph{et~al.}, 1994).
For our Galaxy, Olling and Merrifield (2000) reported
a value of $c / a$ approximately
equal to 0.8 that was obtained, however, using galaxy
parameters (distance to the galactic center and
local galactic rotation speed)
that differ considerably from those recommended
by the IAU; thus, it is unclear whether the method followed
for this derivation yields sensible results. In addition,
van der Marel (2001) 
using the velocity ellipsoid of halo stars to
probe the halo potential of the Galaxy estimated
a lower limit of the ratio ${c / a} >0.4$.
The CDM expectations for the MW were also contested by the study carried
out by Ibata \emph{et~al.}  (2001).
In this study, Ibata \emph{et~al.}
managed to reproduce satisfactorily
enough observations related to cool carbon giant stars in
the galactic halo, assuming  a standard spherical model for the
galactic potential. Thus, they concluded that the galactic dark
matter halo  is most likely almost spherical, at least between the
galactocentric radii of 16 kpc to 60 kpc.
Flat halos with $c/ a <0.7$ were ruled out at very high
confidence levels.

Lastly, lensing data for 20 strong lenses indicated that the mass
distributions are aligned with the luminous galaxy; a 10$^{\circ}$
upper limit on the rms dispersion in the angle between the major axes
of the dark and the luminous components'
distributions has been found (Kochanek, 2001).
This can be interpreted as that light traces the mass, and
this light to mass correspondence is sometimes used as
evidence for that there is no
need for dark matter.

\section{EXCESSIVE SUBSTRUCTURE}

\begin{figure}
\epsfig{file=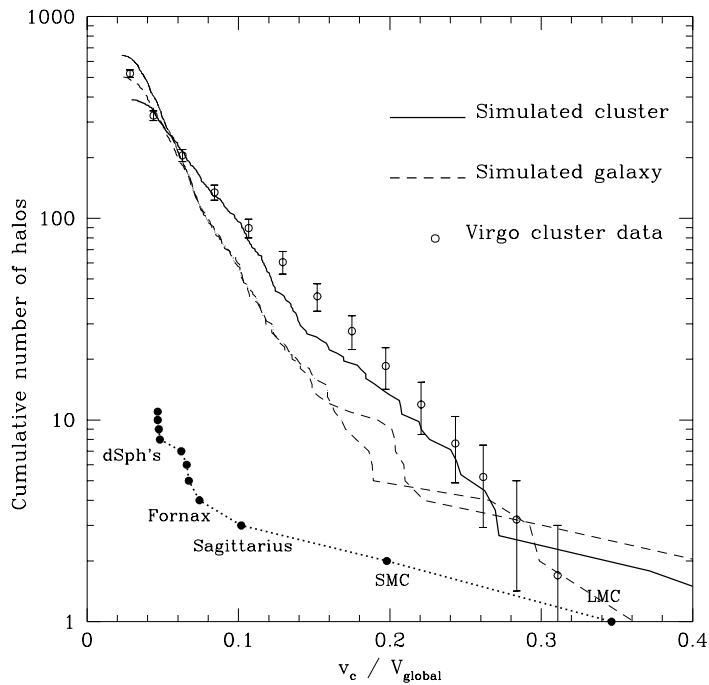, width=9.6cm, height=9.6cm}
\caption{\label{five} [Taken from Moore {\it et al.} (1999a).] 
Abundance of cosmic substructure within the MW,
the Virgo
cluster, and within a simulated cluster and a simulated galaxy. The dotted
curve shows the distribution of satellites within the MW halo and the
open circles with Poisson errors are data for Virgo. The dashed lines
correspond to the substructure in a simulated galaxy today and
4 billion years ago. The solid line corresponds to a cluster mass halo.
The agreement of theory and observation on large scales and the
disagreement on galaxy scales is obvious. First, the simulation
predicts comparable substructure on  large and smaller scales; second,
it predicts excessive substructure on galactic scales compared with
observations.}
\end{figure}

According to the hierarchical clustering scenario, galaxies
are assembled by merging
and accretion of numerous satellites of different sizes and masses
(e.g., Klypin \emph{et~al.}, 1999).
In the frame
of this scenario, smaller galaxies collapsed earlier (when the density
of the universe was higher) and then they participated in the
assembling of the new galaxy.
 This process is not $100 \%$ efficient in destroying
these smaller objects, especially the ones with adequately large central
densities
whose central parts, at least, can survive the merging process and
exist as subhalos within larger halos.
Furthermore, some of the satellites may have been
accreted  by the galaxy at later stages. Both high-resolution N-body
simulations (e.g., Klypin \emph{et~al.}, 1999;
Moore \emph{et~al.}, 1999a; Springel, 2000)
and  semianalytic theory  (Press and Schechter, 1974;
Kauffmann  \emph{et~al.}, 1993) predict
the existence of
 considerable substructure 
 in CDM halos. This substructure has been calculated
 to  amount to of order $10 \%$ of the
halo mass  and to continue down to objects
of unresolved scales of $10^{6}$ M$_{\odot}$
or smaller (Silk, 2001). Furthermore, numerical
simulations have revealed that
the abundance of dark matter subhalos within a galaxy is the same as
found within a scaled
galaxy cluster (e.g., Moore \emph{et~al.}, 1999a; also see Fig.~\ref{five}).

Observation does not seem to verify these CDM predictions and this
discrepancy appears in literature as the {\it satellite catastrophe}
(e.g., Evans, 2001).
The number of predicted
dwarf-galaxy satellites exceeds that
observed around the Milky Way or the Andromeda galaxy by at least
one  order of magnitude (Klypin \emph{et~al.}, 1999;
Moore \emph{et~al.}, 1999a). Klypin \emph{et~al.} (1999)
estimated that a halo the size of
our Galaxy  should have about  50 dark matter halos with circular
velocities $ > 20$  km/sec and mass greater than $3 \times 10^{8}$
M$_{\odot}$
within a $570$ kpc radius.
The same study found that
satellites with circular velocities $10$ km/sec-$20$ km/sec
are approximately
a factor of 5 more than the number of satellites actually observed in the
vicinity of the Milky Way or the Andromeda galaxy.
Moore \emph{et~al.} (1999a) 
found that the virialized extent of the Milky Way halo should
contain about 500 satellites with masses $\geq 10^{8}$ M $_{\odot}$ and
tidally limited sizes $\geq 1$ kpc.
As observed, the Milky Way
contains just 11 satellites within its virial radius  with velocity
$ \geq 10$ km/sec (Mateo, 1998, and references therein).
In the case of the Local
Group, the number of dwarf galaxies is an order of magnitude lower
than predicted
by simulations, with the discrepancy growing towards smaller
masses (Klypin \emph{et~al.}, 1999; Moore \emph{et~al.}, 1999a).
Both the numerical simulations mentioned 
and semianalytic
theory (Press and Schechter, 1974; Kauffmann \emph{et~al.}, 1993) predict
that there should be roughly 1000 dark matter halos within the Local Group,
whereas observations reveal about $40$ (Mateo, 1998). Although
more and more
galaxies are being discovered, most of the new galaxies are very small
and  faint, making it thus  unlikely that too many larger satellites have
been missed. 

Thus, either CDM is incorrect and the (different) nature of dark matter
suppresses the formation of substructure or CDM is correct, but we are
missing something. One of
the first
semianalytic scenarios that appeared in literature in order
to resolve the apparent discrepancy assumes
that the dynamical friction is as efficient as necessary
(highly efficient -- not many clumps could have survived)
so that the
abundance of low-mass satellites  agrees with observations.
Nevertheless, this solution
comes at the expense of the total destruction of larger-mass
satellites;
a dynamical friction capable of destroying a large fraction of low-mass
clumps would also make difficult the survival of any system of the
size of the Magellanic Clouds (Kauffmann \emph{et~al.}, 1993). Other
scenarios use physical
processes that may have operated during the early stages of galaxy formation
and could have resulted in many dark, in the sense of invisible,
satellites (e.g., Navarro and Steinmetz, 1997;
MacLow and Ferrara, 1999; Gnedin, 2000; Somerville, 2002;
Stoehr \emph{et~al.}, 2002). Many of these scenarios,
with the suppression of gas accretion in low-mass halos after the
epoch of reionization the most complete and natural scenario
so far (e.g., Bullock \emph{et~al.}, 2000), 
manage to predict considerably less substructure, in good agreement
with observations.
The basic ingredient of these scenarios
is a mechanism that leads to efficient
mass loss prior to star formation.  However, one can argue
that star formation is a local
process
that occurs in localized inhomogeneities, and the free-fall time is
short because the sound crossing time is short, whereas stripping is
a global process that
requires star formation to first occur (Silk, 2001),
and thus   it is
expected that stars have formed before the gas had been entirely stripped,
and thus the satellites cannot  be completely dark.

Even if there is a mechanism that
inhibits star formation in small
satellites and thus makes them invisible, there
is one more aspect that needs to be examined;
that of a possible problem with respect to spiral  disk formation and
stability (Toth and Ostriker, 1992; Weinberg, 1998;
Moore \emph{et~al.}, 1999a).
In the presence of large amounts of substructure,
the strongly fluctuating potential~\footnote{See footnote 3,
Sec.~\ref{massdist}.}
during clumpy collapses inhibits
disk formation and has been shown to lead to the formation of
elliptical galaxies instead (e.g., Steinmetz and Muller, 1995). But,
old thin disks as well as cold stellar streams
have been observed (e.g., Shang \emph{et~al.}, 1998).
As regards the stability of the disk,
the factor that determines whether the disk is in danger or
not is the amount of energy transferred from the numerous substructure
clumps to the disk. Moore \emph{et~al.} (1999a)
found that
the substructure clumps
are on orbits that take a large fraction of them through the stellar
disk and consequently  the passage of these lumps  will heat the disk
significantly.
They estimated  the energy input
from encounters  in the impulse
approximation -- which however is inadequate
here -- and found that this energy is a significant
fraction of the disk binding energy.
Waker \emph{et~al.} (1996) and Velazquez \emph{et~al.} (1999)
on the other hand,
have shown that disk overheating and stability
become real problems only when interactions with numerous, large satellites
(as for example  the Large Magellanic Cloud) take place.
Since such large
satellites are rare, these two
studies conclude that disk stability does not
impose any constraint on the substructure predicted by CDM N-body simulations. 
More recently,
Font \emph{et~al.}  (2001) concluded as well that there is no conflict
between the existence of numerous satellites and the existence of stellar
disks. Contrary to what Moore \emph{et~al.} (1999a)
concluded,  Font \emph{et~al.} 
find that the orbits of  satellites
in present-day CDM halos very rarely take
them near the disk, where tidal effects become extremely important.
Thus, under the assumption that this was always the case, namely the effects
of substructure at earlier times  were similar to those in the case
of present-day halos, the predicted substructure does not appear
incompatible  with  stellar disks, since there appears to be no
significant issues of  disk overheating and stability.

What sometimes is presented as
another manifestation of excessive substructure in CDM models -- and
on the scales we are concerned with -- is
the failure,  at high confidence level ($ > 99.9 \%$ )
to account for the velocity dispersion distribution
of $z>1.5$ damped Lyman-$\alpha$
systems~\footnote{Systems (clouds)
that go up to column densities $\simeq 10^{21}$ cm$^{-2}$ and
that cause
the Lyman-$\alpha$ forest observed in quasar spectra. They are called
damped because for  column densities higher than $10^{18}$ cm$^{-2}$
the optical depth becomes unity at a point where the line profile is
dominated by the Lorentzian wings of the damped profile rather
than the Gaussian core (see, e.g., Peacock, 2000).} (Prochaska and Wolfe, 1997, 2001).
A theory for these systems must
be able to reproduce both their number density and their velocity
dispersion distribution.
So far, with our current understanding  of these systems,
CDM appears to be unable to reproduce both; even when
it comes close enough, it does so assuming a circular velocity dependence
of the {\it cross section} (gaseous extent) of the systems
that is different from the one obtained
by CDM numerical simulations (Prochaska and Wolfe, 2001, and
references therein).
Note, however, that there are yet a lot to be understood
about the gas dynamics in these systems, and it is
premature to attribute any apparent inconsistency  to
the CDM model rather than to our incomplete
understanding.

Recapitulating, contrary to what was initially believed, the
survival of considerable substructure does not seem to be
in conflict with the existence of disks. With respect to
the substructure abundance, there is either a discrepancy, or
there is no discrepancy and, e.g., numerous dark matter satellites
exist, but have not been detected yet because they have no stars.  
Alternative ways of detecting these
dark matter clumps (e.g., Tasitsiomi and Olinto, 2002) might help
resolve this issue.

\section{THE ANGULAR MOMENTUM OF DISK GALAXIES}

\begin{figure}
\epsfig{file=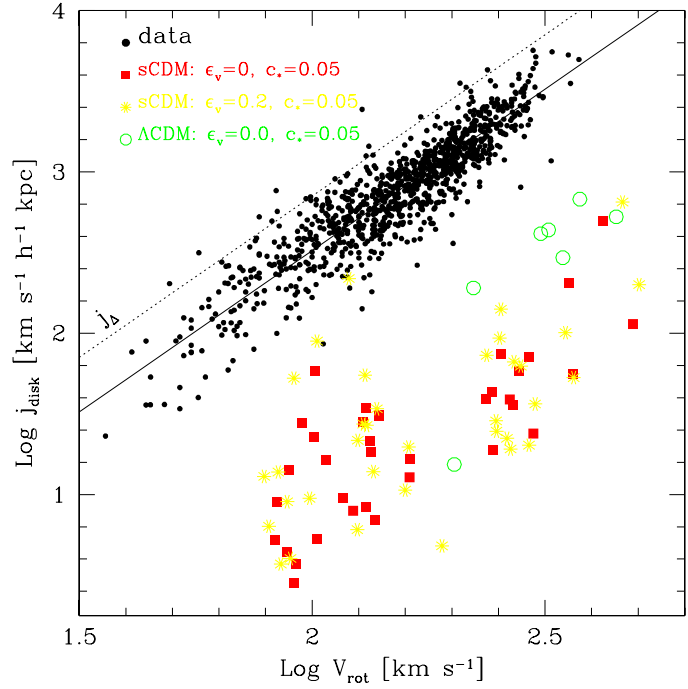, width=9.6cm, height=9.6cm}
\caption{\label{six}
[Taken from Navarro and Steinmetz (2000b).] 
Specific angular momentum against circular velocity of model
galaxies compared with observational data ($c_{*}$ and $\epsilon_{\nu}$
are parameters in the star formation and feedback algorithm used). The
dotted line represents the (halo) velocity-squared scaling
of the halo angular momentum. The solid line
represents the (disk) velocity-squared scaling of the
disk angular momentum.
Note that the angular momenta resulting from
the simulations are at least one order
of magnitude lower than what is observed.}
\end{figure}

Another inconsistency,  between CDM predictions
and observations, that is  closely related to
both the persistence of substructure and
the TFR problems, exists.
It is the one known as the  {\it angular momentum
catastrophe} (e.g., Evans, 2001). It
pertains to the fact that the predicted
angular momentum of disks in spiral galaxies is at least one order
of magnitude less than that observed
(e.g., Navarro \emph{et~al.}, 1995; Navarro and Steinmetz,
1997, 2000a, 2000b;
also see Fig.~\ref{six}).

The reason for this discrepancy between the model and the real
galaxies is that during  galaxy assembling,
clumpiness induces strong angular momentum  transfer
from the  dissipating baryons to
the energy-conserving
dark matter (Navarro and Steinmetz, 1997, 2000b;
Steinmetz and Navarro, 1999). There are two
distinct mechanisms contributing to this angular momentum
exchange. The one is the dynamical friction acting on the orbiting
gas clumps; the other, a global mechanism, 
 is due to the gravitational torques exerted on the orbiting
gas clumps by the non-spherical  dark matter distribution
(e.g., Katz and Gunn, 1991; Navarro and Benz, 1991).
Note that if the specific angular momentum is
conserved  during baryon infall and accretion -- an assumption
that is common in  analytic treatments -- then
the initial angular momentum of a typical protogalaxy when
the halo first collapsed could reproduce the angular momenta
observed [see, e.g., Fall and Efstathiou (1980)].

The baryonic component of simulated disks
retains,
on average, less than $15\%-20\%$ of the specific angular  momentum
of their surrounding
halos (Navarro \emph{et~al.}, 1995; Navarro and Steinmetz, 1997).
The connection
between the angular momentum  and the TFR problem is demonstrated
clearly using the velocity-squared scaling relations that
hold for both the halo and the disk angular
 momentum.~\footnote{ ${j_{disk}/ j_{halo}}
\simeq ({v_{rot}/v_{virial}})^{2}$, with
j denoting the specific angular momentum.
}
These scaling relations
imply that a disk will have retained  about half of the available
angular momentum during assembly, if its rotation speed is
approximately the same as the circular
velocity of its surrounding halo
\footnote{Compare to TFR, Sec.~\ref{tfrb}.} (Navarro
and Steinmetz, 2000b). It has also been
found that the disk
transfers more than $50 \%$ of its original angular momentum
to the halo (between $48 \%$ and $73 \%$) (Katz and Gunn, 1991). The
losses of
angular momentum that take place seem too large for the model to be a
viable
mechanism for making real spiral disks (Navarro and White, 1994).
Most of the
gas  populates extremely low angular momentum orbits, and
only a mass fraction of about $10 \%$ has an angular momentum
comparable to that of
the collisionless dark matter particles (Navarro and Benz, 1991).

Another  equivalent way to show that the angular momenta of the
model disks are deficient, compared with the observed spiral galaxies,
is to compare the
sizes of the former with the luminous radii of the latter.
The equivalence relies
on the fact that the angular momentum losses cause a large contraction
factor to be required for a model disk to reach centrifugal
equilibrium (Navarro and White, 1994; Navarro \emph{et~al.}, 1995).
The scale-lengths of simulated
disks are predicted to be too
small by a factor of $\simeq 5$  compared
with observation (see, e.g., Katz and Gunn, 1991; Steinmetz
and Navarro, 1999).
Again, if angular momentum were conserved during the merging process,
the predicted and the observed sizes
would be in agreement (Fall and Efstathiou, 1980).

To solve this problem we need 
a mechanism that can serve 
as a source of energy that would significantly
reduce the amount of gas that can cool (and delay the cooling)
 and that would prevent the
baryonic component from sinking in the deep cores of the nonlinear clumps
at high redshifts, and from losing a large fraction of their angular
momentum
during subsequent mergers. For this purpose,
for example, SNe feedback (e.g., Thacker and Couchman, 2001),
 or a photoionizing
 UV background (Navarro and Steinmetz, 1997) have been proposed.
The problem is that thus we are led to delayed disk formation. In particular,
giant disks would form relatively late, possibly in conflict  with
the observational data at $z \simeq 1$, as well as with the ages in the
outer parts of nearby disks (Silk, 2001, and references therein).
Moreover,
it is
noteworthy that  disks forming in the presence of a photoionizing
UV background have angular momenta that are even lower than the
angular momenta
of disks forming without including the effects of such a background
(Navarro and Steinmetz, 1997).
Last, but not least,
the loss of angular momentum appears to be a sensitive function of the
numerical resolution used in simulations, becoming more intense with
increasing resolution (Navarro and Steinmetz, 1997).

Nevertheless, the evidence about the existence of a discrepancy should
be considered inconclusive. All this evidence is derived on
the basis of a premature understanding and treatment of the systems
at hand. Thus, there are often problems concerning the ways the CDM
theoretical predictions are obtained.
For example,
until recently a usual assumption was that both dark matter and baryons
have initially similar specific angular momentum
distribution~\footnote{This
was assumed based on the idea that the angular momentum  will be similar
across the halo since it arises from
large scale tidal torques [see,
e.g., Primack (2001), and references therein].}
which -- taking into account the predicted distribution for
the dark matter
(e.g., Bullock \emph{et~al.}, 2001a) -- means
that there will be too much
baryonic matter having a very low  angular momentum to
form the observed, rotationally supported disks. Better
agreement with observation has been recently achieved assuming different
angular momentum distributions for the two components (Vitvitska
\emph{et~al.}, 2001).

\section{EPILOGUE}

Despite its astonishing successes, the  CDM model {\it appears }
to be problematic on galactic scales.
Concluding  with certainty whether the  model simply
{\it appears},
or indeed {\it is} irreparably problematic, is one of the big
challenges of our times.

In order of increasing radicalness,
the opinions appearing in literature regarding the
way the apparent weaknesses of the CDM model should be
handled  can be classified as follows:
\begin{itemize}
\item {\it The CDM model is too compelling
to be wrong.} Thus, to achieve agreement between CDM and
observation we must either add a feature in the initial
power spectrum of inflation, e.g., a tilt in the
spectrum (Alam \emph{et~al.}, 1990; Bullock, 2001a)
that favors structure formation
on large scales while suppresses it on small scales, or  elaborate
on the astrophysics of galaxy formation, e.g.,  to assume the formation
of Super-Massive Black Holes (SMBHs) at the centers of the
dark halos (Gebhardt \emph{et~al.}, 2000;
Menou \emph{et~al.}, 2001; Silk, 2001).
\item {\it There is something wrong with the CDM model.} The way
to solve the problem is by stripping the collisionless or the cold
properties of the traditional CDM, or by considering additional
exotic properties for it; thus a plethora of studies
(e.g.,  Colin \emph{et~al.}, 2000; Kaplinghat
\emph{et~al.}, 2000; Spergel and Steinhardt, 2000; Alam \emph{et~al.}, 2001;
Bode \emph{et~al.}, 2001) assuming particles
that are self-interacting, warm, fluid, annihilating,  etc.,
have been carried out.
\item {\it The CDM model is wrong.} This is the opinion
of the most ardent adversaries not simply of the CDM model, but of
dark matter itself. A solution here is, e.g.,
the Modified Newtonian Dynamics (MOND) scenario
(e.g., McGaugh and de Blok, 1998;
Sanders and Verheijen, 1998).
\end{itemize}

So far, none of these three  alternative~\footnote{But not equivalent,
as becomes clear from their classification with respect to their
radicalness.}
ways of thinking
 has been proven to be the
panacea the problem is in need of. Each one of them  solves
some of the problems either leaving, in the best case, the other problems
unsolved or generating, in the worst case, new problems.

Furthermore, not all  the problems discussed in this
paper are of the same nature.  Thus, there are problems  that
might not only be problems of the CDM model, but also 
of the dark matter hypothesis itself. 
There are problems that appear to be closely related to
the cold nature of the dark matter and its implications,
as  for example the central
halo cusps or  the excessive substructure.
There are problems concerning the ways the CDM theoretical
predictions  are obtained, such as problems with  the assumptions made
or  problems with the
basic tool used to explore the CDM  theoretical
predictions: the simulations. Simulations
are a huge chapter in the CDM debate. Apart from resolution and
convergence issues, and from difficulties in comparing
different simulations, due to different numerical techniques,
cosmological models, etc.,
the really important question is how close simulations are to
reality.  To make realistic predictions of galactic properties
in cosmological theories one must take into account a series of
hydrodynamical phenomena, such as baryonic infall, and must
understand and model important processes such as star formation and
SN feedback. Thus, an important issue 
is to understand whether it is the CDM model or the simplified
treatment of physics in simulations
that is the source of the controversies.
It is also  important to understand 
whether making simulations
be very close to reality -- which  can be extremely
costly -- is a necessity or
not, and whether, instead, we can benefit more 
through  semianalytic treatments.

From what has been already mentioned, it is clear that understanding
and modeling
in the correct way the related
physical processes might resolve,  or might have
already resolved, some of the CDM shortcomings, such 
as the angular momentum catastrophe, the
zero-point of the TFR, or even the satellite catastrophe. 
The  central mass concentration problem
appears to be the most robust discrepancy. What is more, studies
(e.g., Blumenthal \emph{et~al.}, 1986;
Kochanek \emph{et~al.}, 2001)  of the effect
of baryonic infall on the dark matter distribution
conclude that the dark matter is adiabatically compressed by the
cooled baryons during the formation of the central galaxy and is
drawn  towards the center; as a result, it appears so far that
baryonic infall acts towards more centrally concentrated halos
than the ones predicted by collisionless N-body simulations and
thus exacerbates the problem. But, this is not necessarily
the end of the story;
for example, in the case of a barred galaxy,
after including baryonic infall, one must include
the dynamical friction  from
the bar that may act in a way  opposite to the way baryonic
infall acts, and so forth.  

Before concluding with respect to the fatality of the apparent
problems of the CDM model, the problems pertaining to
observations should be noted. Even though
the amount of information that we obtain via observation increases
considerably day-by-day, there are still important accuracy and
resolution issues and quite often, there are issues of correct
interpretation of the observations. Another  big problem
is the fact that observational constraints are strongest just where
theoretical predictions are least trustworthy. 

With respect to the severity of the problems discussed, it seems
that if a problem proves to be fatal for CDM, it will most certainly
be the central mass problem, mainly as it manifests itself
via the rotation curves of LSB galaxies. As already mentioned,
a lot of ideas
have appeared  that have already given, or
have promised to
give, solutions to the other problems, or to the other manifestations of
the central mass problem.
Regarding the central mass problem,
the outlook is not that promising. Nevertheless,
 a few  attempts
have been made. To solve this problem we need
a mechanism that enables heat transport to the inner part of
the halo. This will lead to the puffing up of the central region
and to the flattening of the density cusp. Apart from the small number
of studies and proposed scenarios, the generality of the scenarios proposed
is an additional issue.
For example, Weinberg and Katz (2001)
have shown recently that
a  bar can produce cores in cuspy CDM density profiles within
5 bar orbital times, by means of  angular momentum transfer  from
the bar to the cusp that occurs via an inner-Lindblad-like
resonance. The problem, however, essentially remains
since first, this
scenario  assumes the existence of very large bars at early epochs,
and second, it would be
applicable in the case of barred galaxies and thus not in
the case of the
dark matter dominated dwarf and LSB galaxies,
that have small or nonexistent bars.
A few studies
that might be more general have been carried out and  have promising
results, even though some further investigation of their assumptions
is necessary. For example,
recently  El-Zant \emph{et~al.} (2001)
have found that
provided  the gas is not smoothly, but 
in the form of lumps distributed
initially, the dynamical friction that acts on these lumps dissipates
their orbital energy and deposits it in the dark matter; this energy
was found to be
enough to heat the halo and thus eliminate the cusp. However, issues
such as the creation and survival of these lumps  need
further investigation. A  scenario that is very promising 
with respect to the resolution 
of the central cusp problem
is the one motivated by the observed correlation between the mass of
the SMBH at the center of a galaxy
and the mass of the galaxy
spheroidal (e.g., Gebhardt \emph{et~al.}, 2000). The idea  is
that SMBHs are formed at the center of the dark matter halos. The
merging of the halos is accompanied by the merging of their SMBHs.
The SMBH merger results in the heating of the cusp which thus, becomes
more flat (e.g., Merritt \emph{et~al.}, 2000).

If the central mass problem will not turn out to
be fatal, one can be optimistic;
valid possible solutions will be found (and have been  found already),
at least  for each one
of the problems separately. Eventually, the
real challenge will be  how to find a unique, complete, and consistent
scenario that will complement
and extend the CDM model
starting from ideas and scenarios that were designed to
resolve each one of the discrepancies.
Namely, how to complement the CDM model with the correct processes,
so that the it becomes a theory that can encompass all the related
phenomena.

\section*{Acknowledgments}
I would like to express my gratitude to Sean Carroll for
carefully reading the manuscript, for his useful comments and
remarks, for the stimulating conversations and for
his inspiring encouragement.
I wish to thank Andrey Kravtsov for critically reading the draft version
and for his useful and constructive comments.
I thank Angela Olinto for her encouragement, and Craig Tyler for
reading parts of the initial version. Lastly, I would like
to thank Dimitrios Zisoulis for his endless support, and for being
an inexhaustible source of inspiration.
This work was supported by the National Science Foundation grant
NSF PHY-0114422 at the Center for Cosmological Physics at the University of 
Chicago.

\bibliographystyle{apsrmp}


\newpage

\newpage

\newpage

\end{document}